# Multiple Quantum NMR and Entanglement Dynamics in Dipolar Coupling Spin Systems

**September 5, 2008**

*G. B. Furman[1,2], V.M.Meerovich[1], and V.L.Sokolovsky[1]*

*[1]Department of Physics, Ben Gurion University, Beer Sheva 84105, Israel*

*[2]Ohalo College, Qazrin, 12900, Israel*

## Abstract

We investigate numerically the time dependence of the multiple quantum coherences and entanglement in linear chains up to nine nuclear spins of 1/2 coupled by the dipole-dipole interactions. Two models are considered: (1) a spin chain with nearest-neighbor dipole - dipole interactions; (2) a more realistic model with interactions between all spins. It is shown that the entangled states appear between remote particles which do not interact directly (model 1), while the interaction between all spins (model 2) not always results in entanglement between remote spins.

# Introduction

Entangled states [1-4] are a major resource for quantum computing [5], quantum communication [6], and quantum metrology [7, 8]. This stimulates intensive study of the entanglement properties, including both qualitative and quantitative aspects [3, 4, 9]. Low-dimensional systems, such as spin chains at low temperatures, are good candidates for this study [3]. Correlations between remote parts of a spin system play the predominant role in the dynamics of a spin system. On the one hand, the correlations are directly related to entanglement between different spins [3, 4]. On the other, the correlations lead to the appearance of a multiple-quantum (MQ) coherence [10].

It was found that in two and three spin clusters the measure of the spin pair entanglement, concurrence, almost coincides with the intensity of the second order MQ coherences at low temperatures (T<27 mK) [11, 12]. On the other hand, it was shown in [12], that at higher temperatures the entanglement does not emerge while the MQ coherences of all possible orders exist. In this paper, we study the entanglement and MQ dynamics in linear chains up to nine nuclear spins of 1/2, coupling by dipole-dipole interactions (DDI) at low temperature. Numerical simulation of the dynamics is performed for two models: (1) a spin chain with nearest-neighbor interactions; (2) a more realistic model with DDI between all spins.

## MQ NMR dynamics at low temperature

Consider an MQ NMR experiment on a system of nuclear spins ($s = 1/2$) coupled by the DDI in a strong external magnetic field $\vec{H}_0$ at low temperature. At initial time $t = 0$ the spin system is assumed to be in thermal equilibrium with the lattice and the equilibrium spin density operator $\rho_{eq}$ has the following form

$$\rho_{eq} = \frac{\exp(-\beta H)}{Tr[\exp(-\beta H)]} \qquad (1)$$

where $H$ is the Hamiltonian of the system, $\beta = \frac{\hbar}{kT}$ , $T$ is the Zeeman temperature.

The basic scheme of MQ NMR experiments consists of four distinct periods of time: preparation, evolution, mixing and detection [10]. There are many radiofrequency (RF) pulse sequences exciting MQ coherences during the preparation period. For a dipolar-coupled spin system, the multipulse sequence with an eight-pulse cycle is known to be very efficient [10]. It creates the double-quantum effective Hamiltonian

$$\mathbf{H}_{MQ} = \mathbf{H}^{(2)} + \mathbf{H}^{(-2)}, \qquad (2)$$

where $\mathbf{H}^{(\pm 2)} = -\frac{1}{4}\sum_{j<k} D_{jk} I_j^{\pm} I_k^{\pm}$ , $D_{jk} = \frac{\gamma^2 \hbar}{r_{jk}^3}(1 - 3\cos^2\theta_{jk})$ is the coupling constant between spins $j$ and $k$ , $\gamma$ is the gyromagnetic ratio, $r_{jk}$ is the distance between spins $j$ and $k$ , $\theta_{jk}$ is the angle between the internuclear vector $\vec{r}_{jk}$ and the external magnetic field $\vec{H}_0$ which is directed along the *z*-axis, $I_j^+$ and $I_j^-$ are the raising and lowering operators of spin $j$ .

The density matrix of the spin system, $\rho(\tau)$, at the end of the preparation period is

$$\rho(\tau) = U(\tau)\rho_{eq}U^{+}(\tau), \qquad (3)$$

where $U(\tau) = \exp(-i\tau \mathbf{H}_{MQ})$. Then the evolution period without any pulses follows. The transfer of the information about MQ coherences to the observable magnetization occurs during the mixing period. The resulting signal $S(\tau,t)$ stored as population information reads [10, 13]

$$S(\tau,t) = \sum_{n} e^{-in\Delta\omega t} J_n(\tau) \qquad (4)$$

where $\Delta\omega$ is the RF offset, chosen to be larger than the local dipolar field frequency, $\omega_d$. $J_n(\tau)$ is the spectral intensities of order $n$ [13]

$$J_n(\tau) = Tr[\rho(\tau)\rho_{zn}(\tau)]. \qquad (5)$$

Here $\rho_{zn}(\tau)$ was determined as following: first we calculated $\rho_z(\tau) = U(\tau)I_z U^{+}(\tau)$ ($I_z$ is the *z*-component of the spin angular momentum operator) and then we grouped the terms in $\rho_z(\tau)$ according to their MQ order *n*: $\rho_z(\tau) = \sum_n \rho_{zn}(\tau)$.

## Entanglement of spin pairs

One of the most difficult and at the same time fundamental questions in the entanglement theory, is quantifying entanglement. One of the most natural measures of entanglement of a two-qubit system is the von Neumann entropy [14]:

$$E_F(\rho^{red}) = Tr\left(\rho^{red} \ln \rho^{red}\right) \qquad (6)$$

where $\rho^{red}$ is the reduced density matrix. The analytic expression for $E_F$ is given by

$$E_F(x) = -x\log_2 x - (1-x)\log_2(1-x) \; , \qquad (7)$$

where $x = \frac{1}{2}\left(1+\sqrt{1-C^2}\right)$ and $C$ is the concurrence between two qubits [15]. For maximally entangled states, the concurrence is $C = 1$ while for separable states $C = 0$. The concurrence between the spins $m$ and $n$ is expressed by the formula

$$C_{mn} = \max\left\{0,\; \lambda_{mn} - \sum_{k=1}^{4} \lambda_{mn}^{(k)}\right\} \; , \qquad (8)$$

where $\lambda_{mn} = \max\left\{\lambda_{mn}^{(k)}\right\}$ and $\lambda_{mn}^{(k)}(k = 1,2,3,4)$ are the square roots of the eigenvalues of the product

$$R_{mn}(\tau) = \rho_{mn}^{red}(\tau)\left(\sigma_y \otimes \sigma_y\right)\tilde{\rho}_{mn}^{red}(\tau)\left(\sigma_y \otimes \sigma_y\right). \qquad (10)$$

Here $\sigma_y$ is a Pauli matrix. For $m$ -th and $n$ -th spins, the reduced density matrix $\rho_{mn}^{red}$ defined by $\rho_{mn}^{red}(\tau) = Tr_{mn}\left(\rho(\tau)\right)$ , where $Tr_{mn}(...)$ denotes the trace over the degrees of freedom for all spins except the $m$ -th and $n$ -th ones, $\tilde{\rho}_{mn}^{red}(\tau)$ is the complex conjugation.

# MQ dynamics and evolution of entanglement

We shall study the spin system which is initially in thermodynamic equilibrium state Eq.(1). We restrict ourselves to numerical simulations of MQ NMR and entanglement dynamics for one-dimensional spin systems. Examples of such systems are quasi-one-dimensional hydroxil proton chains, which can be formed using calcium hydroxiapatite $Ca_5(OH)(PO_4)$, and fluorine chains in calcium fluorapatite $Ca_5F(PO_4)_3$ [16].

In our numerical simulations, the dipolar coupling constant of the nearest neighbors is chosen to be $D_{j,j+1} = 1\,s^{-1}$ . We assume also that the angles $\theta_{jk}$ are the same for all pairs of spins and the distances between nearest neighbors $r_{jk}$ are equal. Then the coupling constant of spins $j$ and $k$ is $D_{j,j+1} / |\, j-k\,|^3$ . The numerical simulations of the MQ and entanglement dynamics in the chains up to nine spins are performed using the software based on the MATLAB package.

First we will study the model with nearest neighbors interactions. It was shown [13] that, within limits of this model, the profile of MQ coherences at low temperatures consist of the coherences of the zeroth and second orders only: $J_0$ and $J_{\pm 2}$ (Fig. 1a). The contribution of terms of higher order to MQ coherences is equal to zero because the DDI of the nearest neighbors cannot link up the spins for creation of necessary correlations [17]. Along with evolution of the MQ coherences, we examine the time dependence of the concurrence, $C_{mn}(\tau)$ , between the spins $m$ and $n$ by letting the spin system evolve under the Hamiltonian $H$ . In our calculations, the parameter $\beta\|H\|$ (here $\|...\|$ denotes a norm of the operator) which determines the temperature dependence of intensities of MQ coherences is taken 10. For protons in the external magnetic field $H_0 = 5\,T$ , this value corresponds to the temperature of 1 $mK$ . Fig. 1 shows that, in spite of the fact that only the zeroth and second order coherences are generated, entanglement between remote spins (next-nearest, $C_{k,k+2}$, and next-next nearest, $C_{k,k+3}$ ) is created. This result is evidence that the quantum correlation between remote spins exists. The initial period of evolution is characterized by the creation of entanglement only between the nearest neighbors in the chain (Fig. 1b). Then, entanglement develops between distant spins (Fig. 1c and d). The longer is the distance between spins, the more time the appearance of their entanglement takes. Entanglement between the neighbors appears practically simultaneously in any place of the chain (Figs.1 and 2). The numerical simulation shows that the creation time is independent of the chain length. Usually, direct interactions are considered as the decisive factor required for the entanglement creation [18-20]. Here we observe the generation of entangled states between remote particles which do not interact directly (Fig. 1,c and d). Similar effect was obtained also in a one-dimensional Ising chain [21].

The results of the numerical calculation for the model with DDI between all spins of the chain are presented in Figs. 2 and 3. One can see from Fig. 2a that, in the initial period of the MQ evolution the zeroth and second orders only appear in an eight-spin chain. Then at $D_{j,j+1}\tau \approx 7$ the MQ coherences $J_{\pm 4}$ and $J_{\pm 6}$ appear (Fig. 2a). The similar behavior was obtained also in 6, 7, and 9 spin chains. The concurrences between nearest neighbors grow simultaneously in all places in the chain (Fig. 2b). The concurrence $C_{k,k+2}$ between next-neighbor spins increases more rapidly if the spins are located in the middle part of the chain (compare the curves $C_{13}$ and $C_{35}$ in Fig.2c). One can see from the inset of Fig. 2c that entangled quantum states arise practically at the initial stage of evolution. The main difference of the considered case from the model with only the nearest-neighbor interactions is that, for relatively long spin chains (starting with 7 spins), entanglement between the first spin and the spins located at the middle of the chain disappears, in spite of the direct interaction between these spins. For example, for the 7-spin chain $C_{1,5} = 0$, beginning with $D_{j,j+1}\tau \approx 7$. Similar results are obtained for the 8- and 9- spin chains: $C_{1,4} = C_{1,5} = C_{1,6} = C_{1,7} = 0$ at $D_{j,j+1}\tau > 7$. Surprisingly, that, at the same time, the concurrences $C_{1,N}$ between the ends of the chain, i.e. between the most remote spins, are non-zero (Fig. 3). One more unforeseen peculiarity in the behavior of the concurrence is: the concurrence $C_{1,N}$ between the ends of in an eight spin chain is smaller than in a nine-spin chain (Fig. 3).

## Conclusions

MQ NMR methods can be used to generate and study the dynamics of entanglement. It was shown that entanglement appears between remote spins not interacting directly. An unexpected result that, in the model with direct interactions between all spins, entanglement between the first spin and the spins located at the middle of the chain disappears while the spins at the ends of the chains remain to be entangled. Whereas the intensities of higher order MQ coherences in the many spin chains increase in time, the concurrences of the spin pairs reduce sufficiently faster than the second order coherence intensity. A possible reason of the lack of consistency of the dynamics of MQ coherences and entanglement can be the fact that the coherence is a collective phenomenon while the concurrence, as a measure of entanglement, is determined for a definite pair of the spins.

Figure Captions

Fig. 1 (Color online) (a) Time dependence of the intensities of the MQ coherences of the zeroth $J_0$ (black solid line) and second $(J_{+2} + J_{-2})$ (red dash line) orders in a linear chain of eight spins coupled by the nearest neighbor DDI.

Evolution of concurrences for various pairs of spins: (b) nearest spins, $C_{12}$ (black solid line); $C_{23}$ (red dash line); $C_{34}$ (green dot line); $C_{45}$ (blue dash-dot line)

(c) next nearest spins: $C_{13}$ ( black solid line); $C_{24}$ (red dash line); $C_{35}$ (green dot line)

(d) next-next-nearest spins: $C_{14}$ (black solid line); $C_{25}$ (red dash line); $C_{36}$ (green dot line)

Fig.2 (Color online) (a) Time dependence of the intensities of the MQ coherences of the zeroth $J_0$ (black solid line), the second $(J_{+2} + J_{-2})$ (red dash line), the forth $(J_{+4} + J_{-4})$ (green dot line), and the sixth $(J_{+6} + J_{-6})$ (blue dash-dot line) orders in a linear chain of eight spins coupled by the DDI.

Evolution of concurrences for various pairs of spins: (b) nearest spins, $C_{12}$ (black solid line); $C_{23}$ (red dash line); $C_{34}$ (green dot line); $C_{45}$ (blue dash-dot line)

(c) next nearest spins: $C_{13}$ ( black solid line); $C_{24}$ (red dash line); $C_{35}$ (green dot line)

The inset shows that the intensities of coherence of the next nearest spins are generated just with beginning of the MQ NMR irradiation.

Fig. 3 (Color online) Time dependence of entanglement between the end spins of nine- (black solid line), eight- (red dash line), seven- (green dot line), and six- (blue dash-dot line) spin chains.

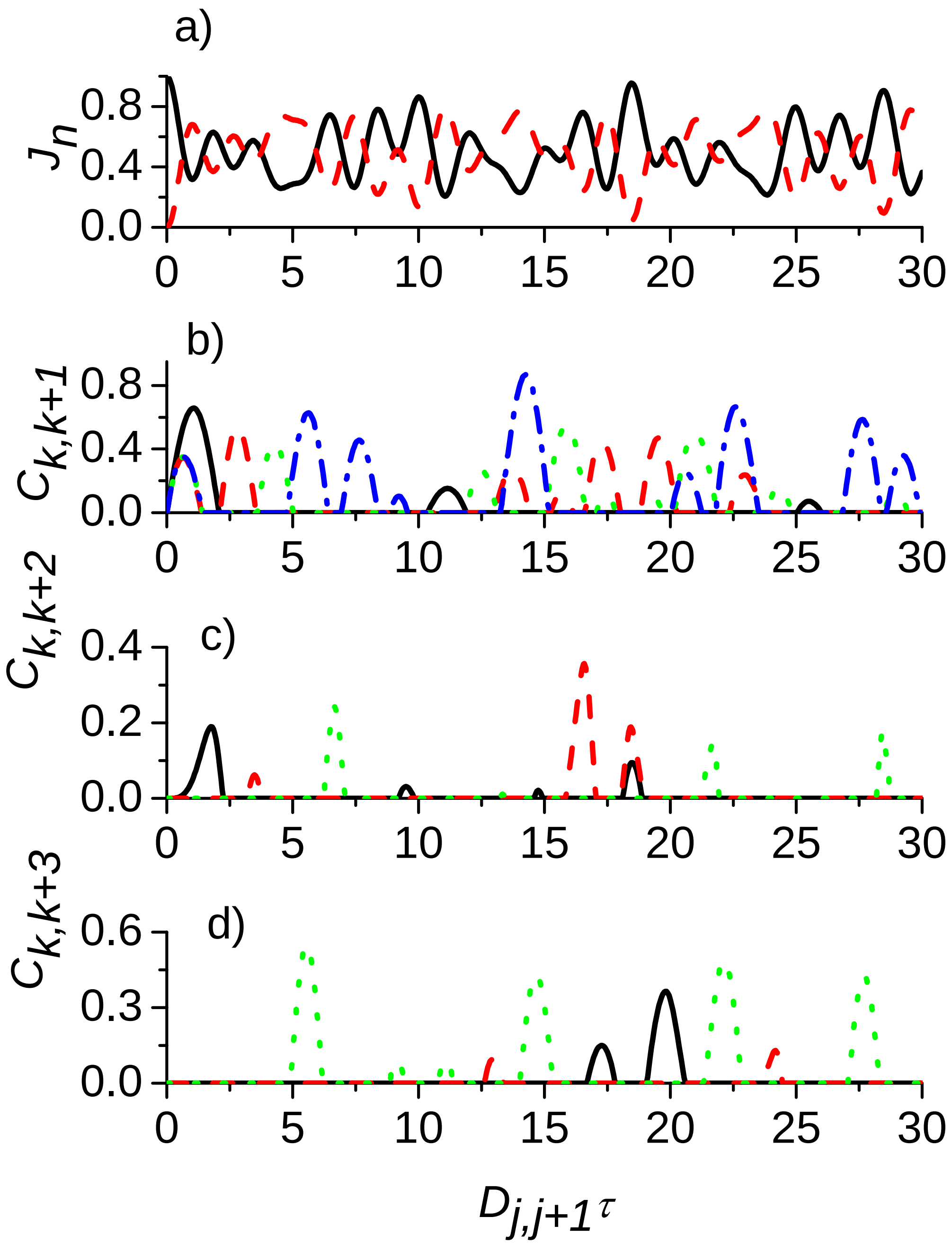

Fig.1

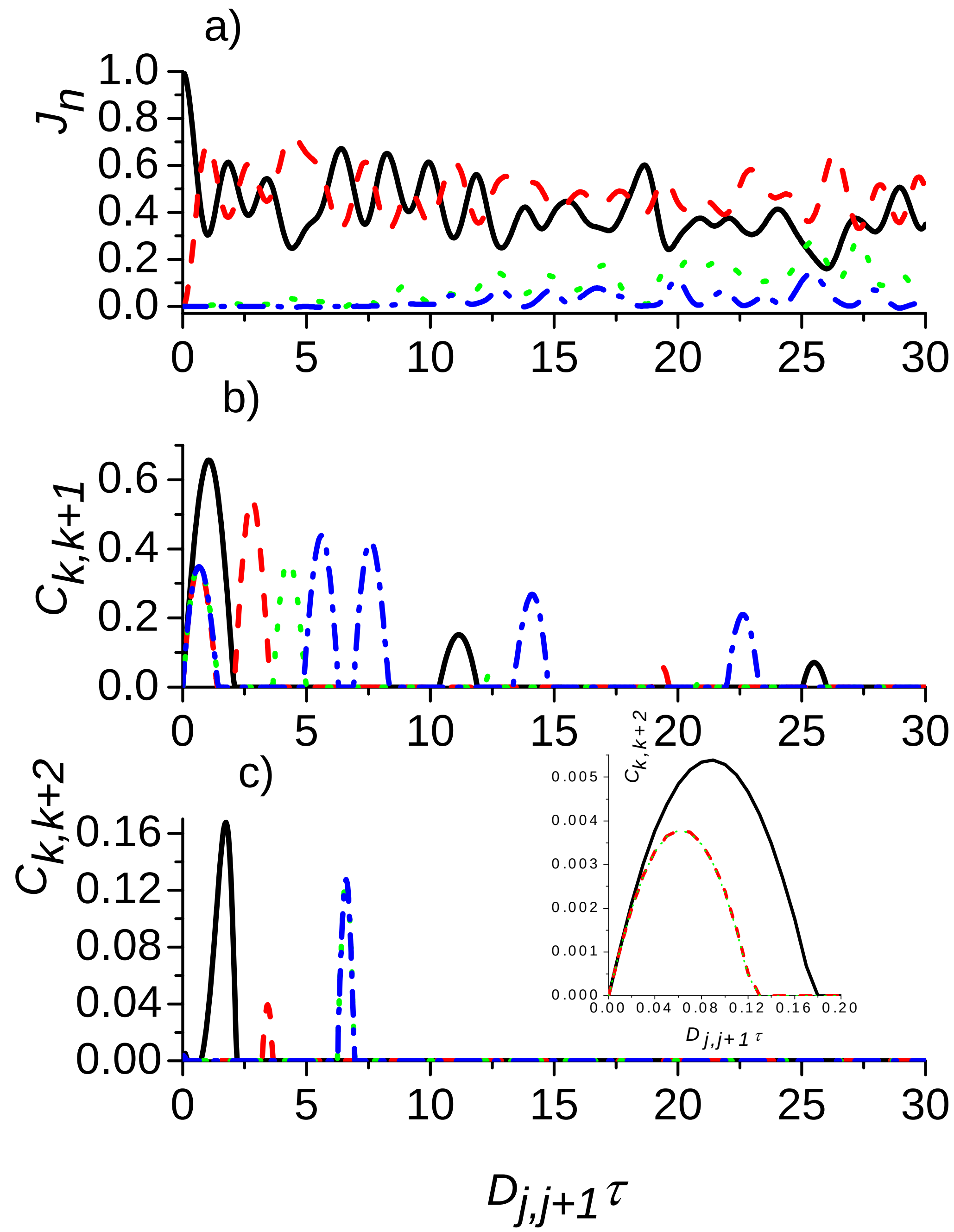

Fig. 2

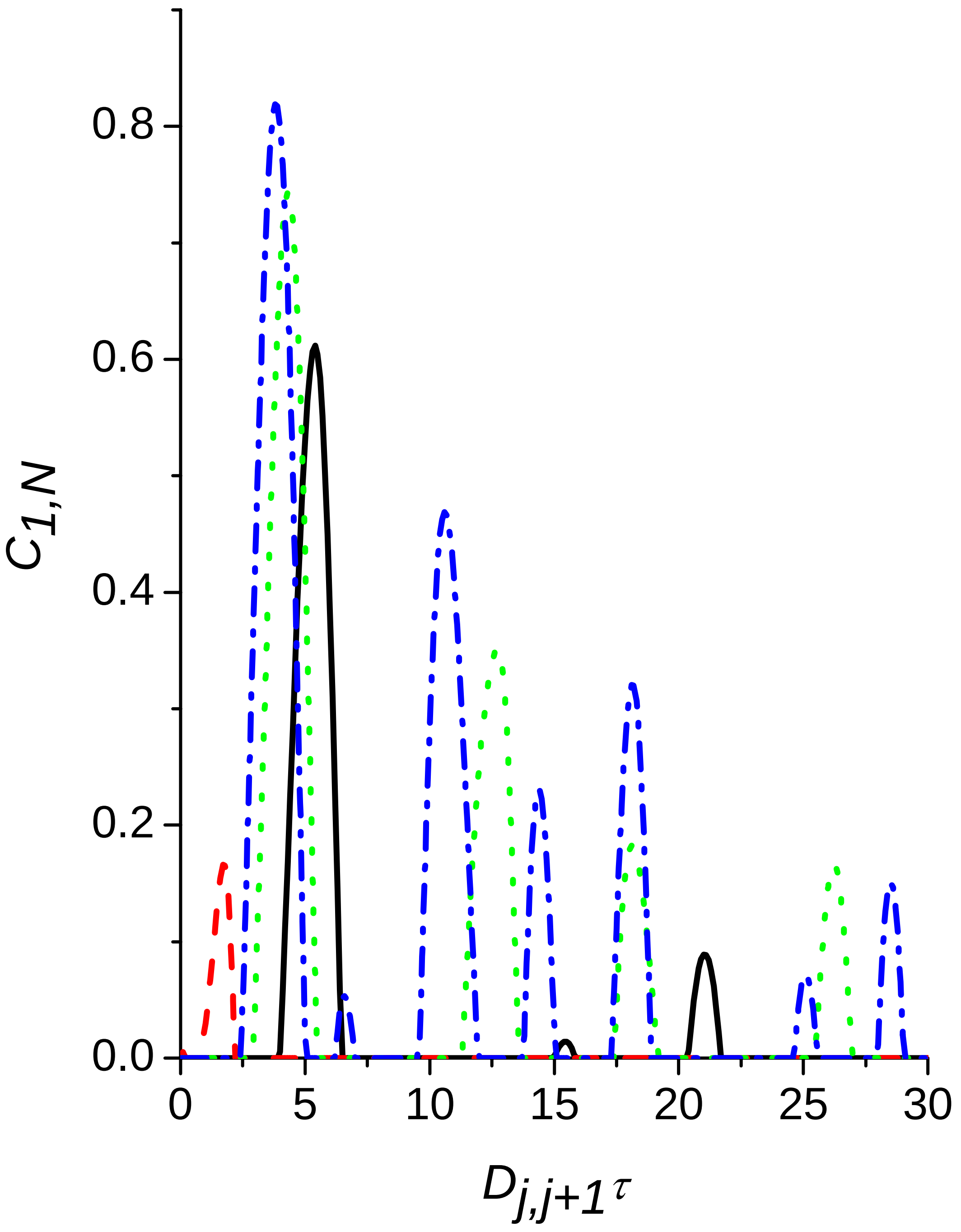

Fig. 3